\documentclass[useAMS,usenatbib]{mn2e}

\usepackage{psfig,latexsym,graphicx,natbib,aas_macros} 

 \usepackage{times}
 \usepackage{lscape}

\bibpunct{(}{)}{;}{a}{}{,}

\pssilent

\title{ The quasi-resonant variability of the massive  LMC system BI 108}

\author[ Ko{\l}aczkowski et al.]
  {   Z. Ko{\l}aczkowski$^{1,2}$ \thanks{E-mail: rmennick@astro-udec.cl. Based on ESO Proposal 382.D-0311},  
  R.E. Mennickent$^{1}$,  
 T. Rivinius$^{3}$,   G. Pietrzy\'nski$^{1, 4}$\\   
  $^1$Universidad de Concepci\'on, Departamento de Astronom\'{\i}a,
      Casilla 160-C, Concepci\'on, Chile\\
  $^{2}$  Instytut Astronomiczny Uniwersytetu Wroclawskiego, Kopernika 11, 51-622 Wroclaw, Poland
 \\
  $^{3}$ ESO, European Organization for Astronomical Research in the Southern Hemisphere, Chile \\
  $^{4}$ Warsaw University Observatory, Al. Ujazdowskie 4, 00-478 Warszawa, Poland.}
\date{}

\pagerange{\pageref{firstpage}--\pageref{lastpage}} \pubyear{2008}

\def\LaTeX{L\kern-.36em\raise.3ex\hbox{a}\kern-.15em
    T\kern-.1667em\lower.7ex\hbox{E}\kern-.125emX}

\begin{document}

\label{firstpage}

\maketitle

\begin{abstract} 
The early B supergiant LMC star BI\,108 is photometrically variable with
a unique light curve; two strong periods are present in an almost precise 3:2
resonance.
We collected spectroscopic data at VLT/UVES, sampling the
supercycle of 10.733 days in ten epochs. We find spectral signatures for a SB2
system consisting of two massive B1 supergiants orbiting at the orbital period
of 5.366 days.  The shorter periodicity resembles the light curve of an
eclipsing binary with periodicity 3.578 days that is not detected in the
 data.  
We  discuss
possible causes for the short periodicity
and  conclude that the 
quadruple system is the more plausible hypothesis. 
\end{abstract}


\section{Introduction}
The object LMC SC9-127519 (OGLE 051343.14-691837) was discovered as a variable
star with unique photometric properties during our multiperiodicity search in
the OGLE-II database \citep{2005AcA....55...43S}.  Independently this object
was announced as two binary systems in 3:2 resonance by
\citet{2008IBVS.5868....1O}.  With the coordinates, the magnitude ($V=13.3$)
and the color ($B-V=-0.142$), the star was identified as BI\,108 in the list
of Brunet et al. (1975), for which they give a spectral type of
b1-2\footnote{Spectral classes are written in lower-case letters when they have been determined from multicolor photometry.}.
\citet{1999MNRAS.306..279S} list b1:II: for the same star.

BI\,108, in fact a binary { or multiple} system rather than a single star,
is likely one of the brightest members of a young open cluster; NGC\,1881
\citep[$\log t <$ 6.7,][]{2000AcA....50..337P}.
Although the resonance is not absolutely perfect, such a near resonant and
long-term stable period relation is very rare.
In order to investigate the photometric behavior of BI\,108
all available photometric data were used, as well as a set of 20 UVES
observations to determine the spectroscopic behavior.  Brief reports of our
advances in the understanding of BI\,108 were presented in recent conferences
\citep{2010ASPC..435..403K,2011IAUS..272..541R}.



\section{Spectroscopic observations and archival data}

We secured 20 spectra of BI\,108 between October 1, 2008 and January 3, 2009,
with the ESO VLT/UVES. The instrument was used in the 437/760 setting,
covering most important He{\sc i} and Balmer lines and several other
transitions of particular interest for hot stars. At a total of ten epochs two
subexposures were taken with $t_{\rm exp}=1480$\,sec each, with a slit width
of 0.8\,arcsec, giving an effective resolving power $\lambda/\Delta\lambda$ of
about 60\,000. The resulting typical $S/N$ is between 90 and 100 for both arms
of the instrument.

The UVES standard pipeline ran smoothly on our spectra providing wavelength
calibrated 1-D spectra virtually free of instrumental
artifacts. The spectra were 
continuum normalized and heliocentric
corrected. The observing log is given in Table~\ref{tab_specobs}.

Apart from OGLE-II, data were extracted as well from the MACHO (MACHO object ID
79.5378.25) and the OGLE-III database \citep[LMC
  111.2.62038]{2008AcA....58...69U}.  The MACHO data span seven and a half
years (HJD\,2\,448\,826.14 to 2\,451\,541.96), the OGLE-II data almost four
years (HJD\,2\,450\,457.76 to 2\,451\,872.75), and the OGLE-III data over
almost eight years (HJD\,2\,452\,167.64 to 2\,454\,951.52).

The photometric data of all three databases were joined for the analysis. Since
the bands are slightly different, this was done the following way: for the time
series analysis presented in Sect.~\ref{sec_TSA} 
the mean magnitude was subtracted from each dataset. For the MACHO data, only
observations where simultaneous blue and red points exist were accepted. The
two measurements were averaged to reduce the individual uncertainty. Note that
this combined dataset was {\em only} used to determine the periods, not for
any other analysis.  For the actual disentangling of the light curve we chose
the OGLE-II data set only, in order to avoid problems due to the different
filter bands of each dataset.


\section{Photometric properties of the system}

\begin{figure*}
\scalebox{1}[1]{\includegraphics[angle=0,width=18cm]{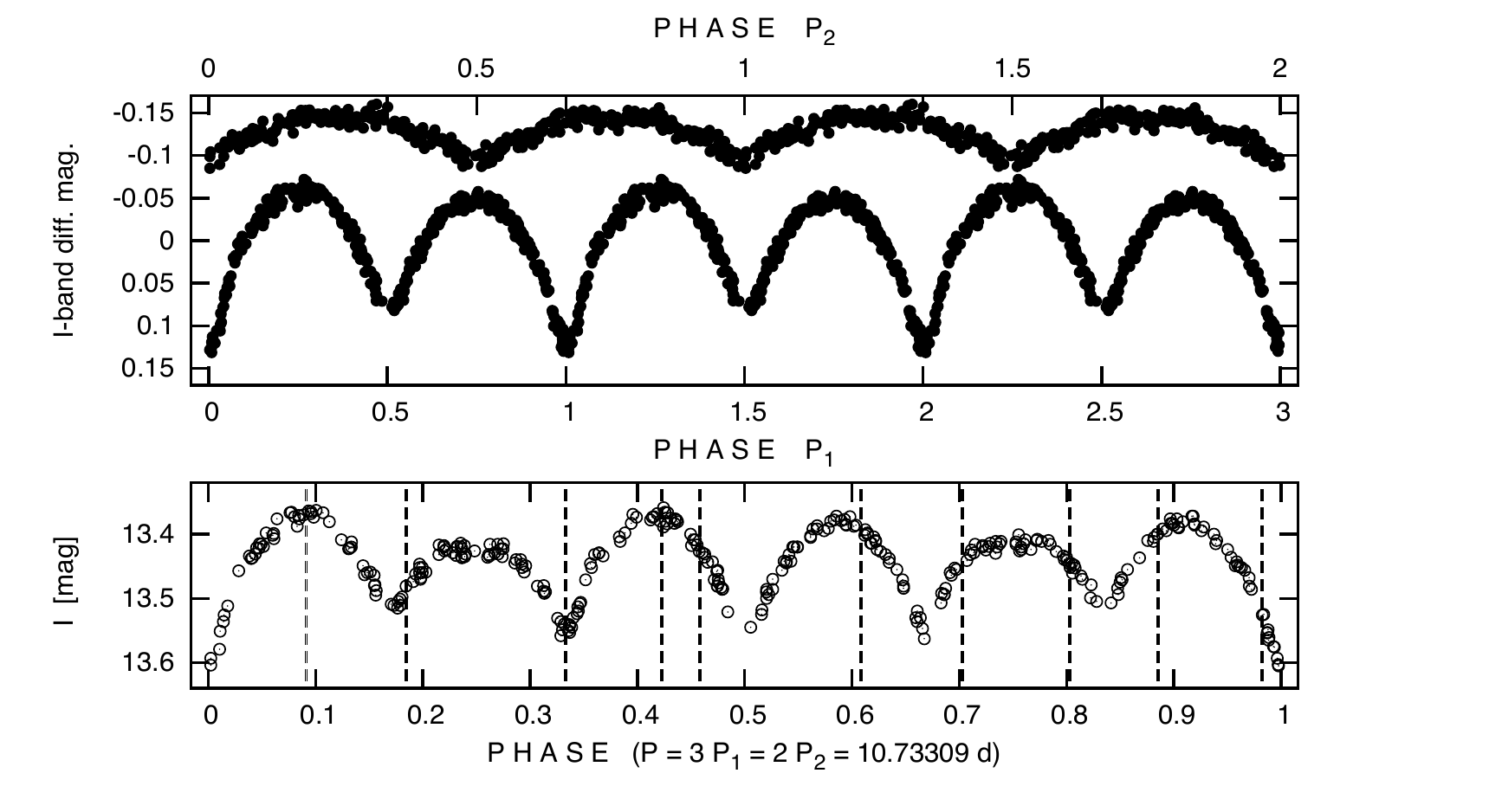}}
\caption{Upper panel: The OGLE-II I-band light curves for both periods of the
  BI\,108 system (upper panel). For each period the contribution of the
  respective other one has been removed.  Lower panel: The original light
  curve folded with the superperiod. In all cases HJD$_0 = 2\,451\,163.8915$. Dashed
  lines indicate phases of the spectral observations. }
  \label{kolaczkowski_fig1}
\end{figure*}

\subsection{Time series analysis and disentangling}\label{sec_TSA}
The time series analysis returns two very strong periods, given in
Table~\ref{tab_alldataTSA}. For reasons that will be shown later, the longer 
one will be called the orbital period ($P_{2}$) and the shorter one the resonant
period ($P_{1}$). From the light curve alone, however, this is by no means obvious. For
none of the two periods is the associated variation  sinusoidal in shape, for
both the minima are more pronounced and sharper than the maximum associated
variations.  In fact both variations are of double wave nature.

\begin{table}
\centering
 \caption{\label{tab_specobs}Summary of spectroscopic observations discussed
   in this paper. We adopted HJD=2\,451\,163.8915 as epoch for the deepest
   minimum. All spectra are 1480\,s exposures, the dates refer to the beginning
   of the exposure }
 \begin{tabular}{@{}ccccc@{}}
 \hline 
Spectrum& HJD &superphase &phase ($P_{1}$) & phase ($P_{2}$) \\
   (label) & (-2\,450\,000) & 10.73\,d & 3.58\,d &5.37\,d\\
\hline   
BI108-01   &  4\,834.63285   &    0.0031 & 0.9318  & 0.0077 \\
BI108-02   &  4\,834.65057   &    0.0047 & 0.9368  & 0.0110 \\
BI108-03   &  4\,742.76051   &    0.4434 & 0.2546  & 0.8882 \\
BI108-04   &  4\,742.77823   &    0.4450 & 0.2595  & 0.8915 \\
BI108-05   &  4\,746.83997   &    0.8234 & 0.3947  & 0.6483 \\
BI108-06   &  4\,746.85768   &    0.8251 & 0.3997  & 0.6516 \\
BI108-07   &  4\,803.60272   &    0.1120 & 0.2593  & 0.2255 \\
BI108-08   &  4\,803.62044   &    0.1137 & 0.2642  & 0.2288 \\
BI108-09   &  4\,761.66650   &    0.2048 & 0.5386  & 0.4111 \\
BI108-10   &  4\,761.68422   &    0.2065 & 0.5435  & 0.4144 \\
BI108-11   &  4\,796.80764   &    0.4789 & 0.3601  & 0.9594 \\
BI108-12   &  4\,796.82536   &    0.4806 & 0.3651  & 0.9627 \\
BI108-13   &  4\,744.75145   &    0.6289 & 0.8110  & 0.2592 \\
BI108-14   &  4\,744.76917   &    0.6305 & 0.8160  & 0.2625 \\
BI108-15   &  4\,741.79717   &    0.3536 & 0.9853  & 0.7087 \\
BI108-16   &  4\,741.81489   &    0.3553 & 0.9902  & 0.7120 \\
BI108-17   &  4\,745.76182   &    0.7230 & 0.0934  & 0.4475 \\
BI108-18   &  4\,745.77953   &    0.7246 & 0.0983  & 0.4508 \\
BI108-19   &  4\,747.72260   &    0.9057 & 0.6414  & 0.8128 \\
BI108-20   &  4\,747.74031   &    0.9073 & 0.6464  & 0.8161 \\
\hline
\end{tabular}
\end{table}


The stronger variation is associated with the shorter period, $P_{1}=
3.577958$\,d, with a peak-to-peak amplitude of about 0.18 mag. From the shape
alone it is tempting to attribute it to an eclipsing binary nature
(Fig.~\ref{kolaczkowski_fig1}, lower curve in the upper panel), however, we will show later 
that at least spectroscopy shows no sign of any binary with that period.

The longer period shows weaker variations, with about 0.05 mag peak-to-peak, and has the
value of $P_{2}=5.366520$\,d (Fig.~\ref{kolaczkowski_fig1}, upper curve
in the upper panel). In the case of $P_{2}$ the two half waves are
indistinguishable in photometry, however, and the double wave nature is shown
by spectroscopy.

The most interesting feature of both periods is that they have extremely
similar superperiods, i.e.\ they are almost fully resonant with a ratio of
$3:2$, or $2\times P_{2}= 3\times P_{1}=10.733$\,d. 
Using the law of propagation of uncorrelated errors we find $P_{2}/P_{1}= 1.499883 \pm 0.000006$.  This differs from 3/2 = 1.500000 by 0.000117 $\pm$ 0.000006, i.e. at the 117/6 = 19.5 sigma level. The
difference between the two superperiods is significant but so tiny, that we will
come back to it in the discussion,
but ignore it for the rest of this work. Instead, we use our OGLE-II
 superperiod of $10.73309$\,d \citep{2011IAUS..272..541R} to analyze variability with the
 supercycle. The combined light curve phased with
that superperiod is shown in the lower panel of Fig.~\ref{kolaczkowski_fig1}.
The minima of the combined curve are governed by the
stronger variation with $P_{1}$, while $P_{2}$ only modulates
these minima, { i.e.\ making them more or less deep.}  We choose the date
of a deepest minimum as epoch, so that the ephemeris for the superperiod
is
\[
T_{\rm min light, super} = {\rm HJD}\,2\,451\,163.8915 + E \times 10.73309
\]
and the ephemerides for orbital and resonant period are 
\[
T_{\rm min light, 2} = {\rm HJD}\,2\,451\,163.8915 + E \times 5.366520
\]
and
\[
T_{\rm min light, 1} = {\rm HJD}\,2\,451\,163.8915 + E \times 3.577958
\]
respectively.

\begin{table}
\centering
\caption[]{\label{tab_alldataTSA}Results of the time series analysis on the
  combined photometric dataset over 16 years.  Amplitudes are MACHO\,b,
  MACHO\,r, and OGLE-II I-band, respectively.  The two superperiods,
  3$\times P_{1}$ and 2 $\times P_{2}$, differ by
  a tiny, yet significant fraction.}
\begin{tabular}{rlll}
&\multicolumn{1}{c}{period} & \multicolumn{1}{c}{superperiod} & 
\multicolumn{1}{c}{amplitude} \\
&\multicolumn{1}{c}{[d]} & \multicolumn{1}{c}{[d]} & 
\multicolumn{1}{c}{[mag]} \\
\hline
$P_{1}$&$3.577958(14)$&  $10.73388(4)$ &0.172, 0.171, 0.178 \\
$P_{2}$&$5.366520(8)$&  $10.733040(16)$  &0.050, 0.047, 0.054 \\
\hline
\end{tabular}
\end{table}

\subsection{Stellar parameters}\label{sec_stelparam}

Using $E(B-V) \approx 0.156$ \citep[for NGC\,1881,][]{2000AcA....50..337P} and
the interstellar extinction law derived by \citet{1998ApJ...500..525S}, we
calculated $(B-V)_0 = -0.30$ \citep[indicating
  b0\,II/III,][]{1970A&A.....4..234F}, $(V-K)_0 = -0.95, (J-K)_0 = -0.28$ and
$(H-K)_0 = -0.18$\,mag. The corresponding figures for a b1\,I\,b star are
$-0.19$\,mag, $-0.55$, $-0.09$ and $-0.03$\,mag, respectively
\citep{1970A&A.....4..234F, 1983A&A...128...84K}.  Although variability could
affect some of these measurements (though not the MACHO colors, which are
taken simultaneously), it is difficult to avoid the conclusion that the
infrared colors are bluer than those of a typical b1 (or even b0 or late o
type) supergiant.

  If we use $V= 13.3$ for the whole system, and assume two equal stellar
components, we get $V= 14.05$ for a single star.  Using the distance modulus
for the LMC of 18.5\,mag \citep{2009ApJ...697..862P}, and the NGC\,1881
reddening $E(B-V)= 0.156$ (Pietrzy\'nski \& Udalski 2000), we obtain
$M_{V} = -4.96$. Using a bolometric correction of $BC = -2.70$ for a star with
$T_{\rm eff} = 28\,000\,{\rm K}, \log\,g = 3.0, v_{\rm turb}=
10$\,km\,s$^{-1}$ and $Z/Z_{\sun} = 0.5$ \citep{2007ApJS..169...83L}, as
representative for BI\,108, we get an absolute bolometric magnitude of $M_{\rm
  bol} = -7.66$.  From the evolutionary tracks for $Z= 0.004$ stars by
\citet{2001A&A...373..555M} this luminosity implies a total mass for the
system of about 40\,$M_{\odot}$. The same exercise with four stars gives 
a total mass of about 60\,$M_{\odot}$.

The MACHO color term, $B_{\rm M}-R_{\rm M}$, does not show any periodic
signature, except a signal with a period of 23h56m, i.e.\ clearly related to
Earth's rotation, and apart from this no other variability at all above the
statistical error, which is 0.01\,mag. 

  Photometric data 
  were retrieved from 2MASS and Spitzer
  observational databases and converted to fluxes using the virtual
  observatory tool VOSA\footnote{http://www.laeff.inta.es/svo/theory/vosa/}
  \citep{2008A&A...492..277B}, and compared to published spectral energy
  distributions.  There is no evidence for any IR excess in BI\,108, i.e.\ a
  significant flux contribution from a cooler component in the visual range is
  firmly excluded.

\section{Spectroscopic variations}

\subsection{Spectral classification of the components}
In the spectra the system obviously has SB2 nature and contains two similar
components with spectral features of B1\,Ib-II stars.  Although both spectra
are virtually identical in terms of lines present and $v \sin i$, in the
combined light one component produces, on average, slightly weaker lines (and
as will be seen from the RV curve is as well less massive). In the following,
this star as being the secondary of the current configuration, will be denoted
star\,\#2, while the other one will be star\,\#1.

At quadratures the spectral lines are almost separated and can easily be
compared to synthetic and other template spectra.  A comparison with the
B1Ib-II galactic star HD\,54\,764 gives an almost perfect match between line
profiles (even in $v\,sin\,i = 125$\,km\,s$^{-1}$). 

The line ratio He{\sc I}\,4387/O{\sc II}\,4416
is a useful luminosity
indicator.\footnote{http://nedwww.ipac.caltech.edu/level5/Gray/frames.html}
The average value is $0.99 \pm 0.09$ for star\,\#1 and $1.1 \pm 0.3$ for
star-\,\#2, in agreement with a luminosity class I/II derived from the visual
classification above.

In spite of what is said above about one star on average being brighter  than
the other, in some of the quadrature phases the two spectra appear of almost
equal strength, i.e.\ the relative contributions from each star to the
equivalent width (EW) are variable. This behavior is discussed in detail
below, here it is sufficient to say that the spectral types, understood as
defined by spectral line ratios, for instance He{\sc i}\,4471/Mg{\sc
  ii}\,4481, do not change significantly with the variable EW.

\begin{figure}
\scalebox{1}[1]{\includegraphics[angle=-90,width=8.5cm]{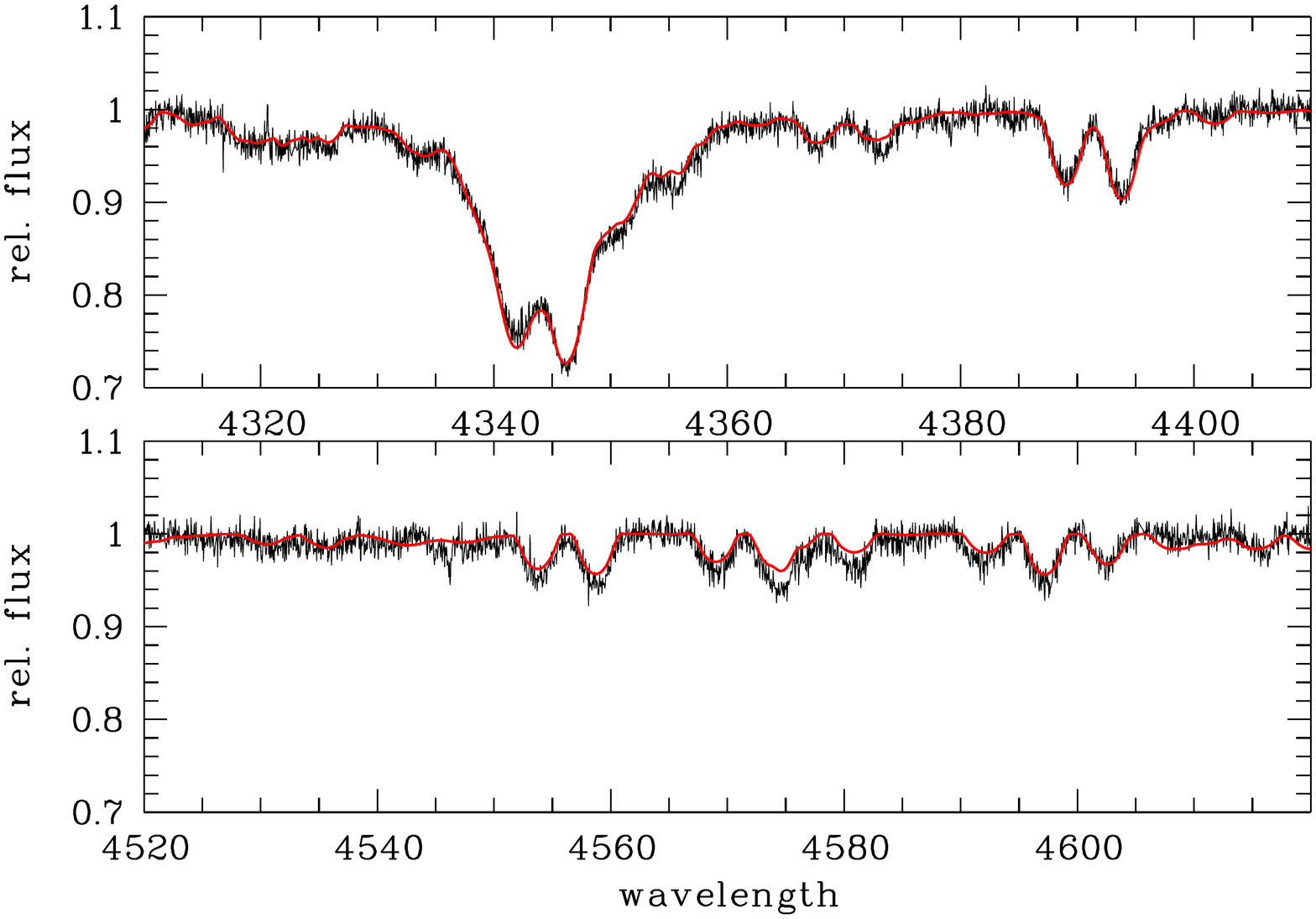}}
\caption{The combined  spectrum at $\Phi_{2}$= 0.81 along with the best fit whose parameters are given in Table 3.} 
  \label{specfit}
\end{figure}

In order to obtain individual stellar parameters for both components, the
combined spectrum was modeled using the grid of synthetic spectra published by
\citet{2003ApJS..146..417L,2007ApJS..169...83L}\footnote{http://nova.astro.umd.edu/Tlusty2002/tlusty-frames-models.html}.
A multi-parameter fit of those models was done to the UVES data taken at good
separation of both components (HJD\,2\,454\,747, $\phi_{2}=0.81$). The
fit parameters were effective temperature, surface gravity (in the range
$T_{\rm eff}$ 25\,kK -- 30\,kK, $\log g$ 3 -- 4), radial velocities and $v\sin
i$ for each component, as well as the relative flux contribution of secondary
component, fixed as unity for the primary (see Table~\ref{tab_spectrofit}).
The calculations were done for models with the average LMC metallicities and
microturbulence velocities $2$ and $10$\,km\,s$^{-1}$.  While the hydrogen and
helium lines are rather insensitive, the higher microturbulence gives the
better fit for the metal lines (Fig.~\ref{specfit}).

\begin{table}
\centering
\caption[]{\label{tab_spectrofit}Results to the fit of the combined spectrum at $\Phi_{2}$= 0.81.}
\begin{tabular}{lll}
Parameter &\multicolumn{1}{c}{star\,\#1} & \multicolumn{1}{c}{star\,\#2}\\
\hline
$T_{\rm eff}$ [K] &  27\,500 $\pm$ 250 &  27\,500 $\pm$ 250  \\
$\log g$        &  3.75 $\pm$ 0.05 & 3.5 $\pm$ 0.05 \\
$v \sin i$  [km\,s$^{-1}$]& 116 $\pm$ 5 & 116 $\pm$ 5 \\
$v_{\rm rad}$ [km\,s$^{-1}$]& 401 $\pm$ 2 & 75 $\pm$ 2\\
relat. flux contrib.   & 1.0 $\pm$ 0.02 & 0.88 $\pm$ 0.02 \\
\hline
\end{tabular}
\end{table}

 \begin{table}
\centering
 \caption{\label{tab_korelres}System properties of BI\,108 derived from the
   spectroscopic data.  }
 \begin{tabular}{lclc}
 \hline
Property &Value &Property &value \\
\hline       
$q$ & 0.75            &$M_1 \sin^{3}\,i$ &19 $M_{\sun}$ \\
 $e$ &0.08            &$M_2 \sin^{3}\,i$ &14.5 $M_{\sun}$ \\
$\omega$ & 93  & $M  \sin^{3}\,i$ &33.5 $M_{\sun}$ \\
$a \sin\,i $ & 41.5 $R_{\sun}$       \\
\hline
\end{tabular}
\end{table}

\subsection{Time series analysis and disentangling}

We measured RVs for both components by finding line centers with deblending
algorithms, as well by cross correlation of selected spectral regions with
templates, but by far the best result was obtained by Fourier disentangling the
spectra with the virtual observatory program
KOREL\/\footnote{http://www.asu.cas.cz/~had/korel.html}
\citep{1995A&AS..114..393H,2010ASPC..435...71S}. Although robust in spectral disentangling and binary parameter determination, KOREL does not provide the zero point of velocity neither meaningful errors for the system parameters.

\begin{figure}
\includegraphics[viewport=55 41 572 775,angle=-90,width=8cm]{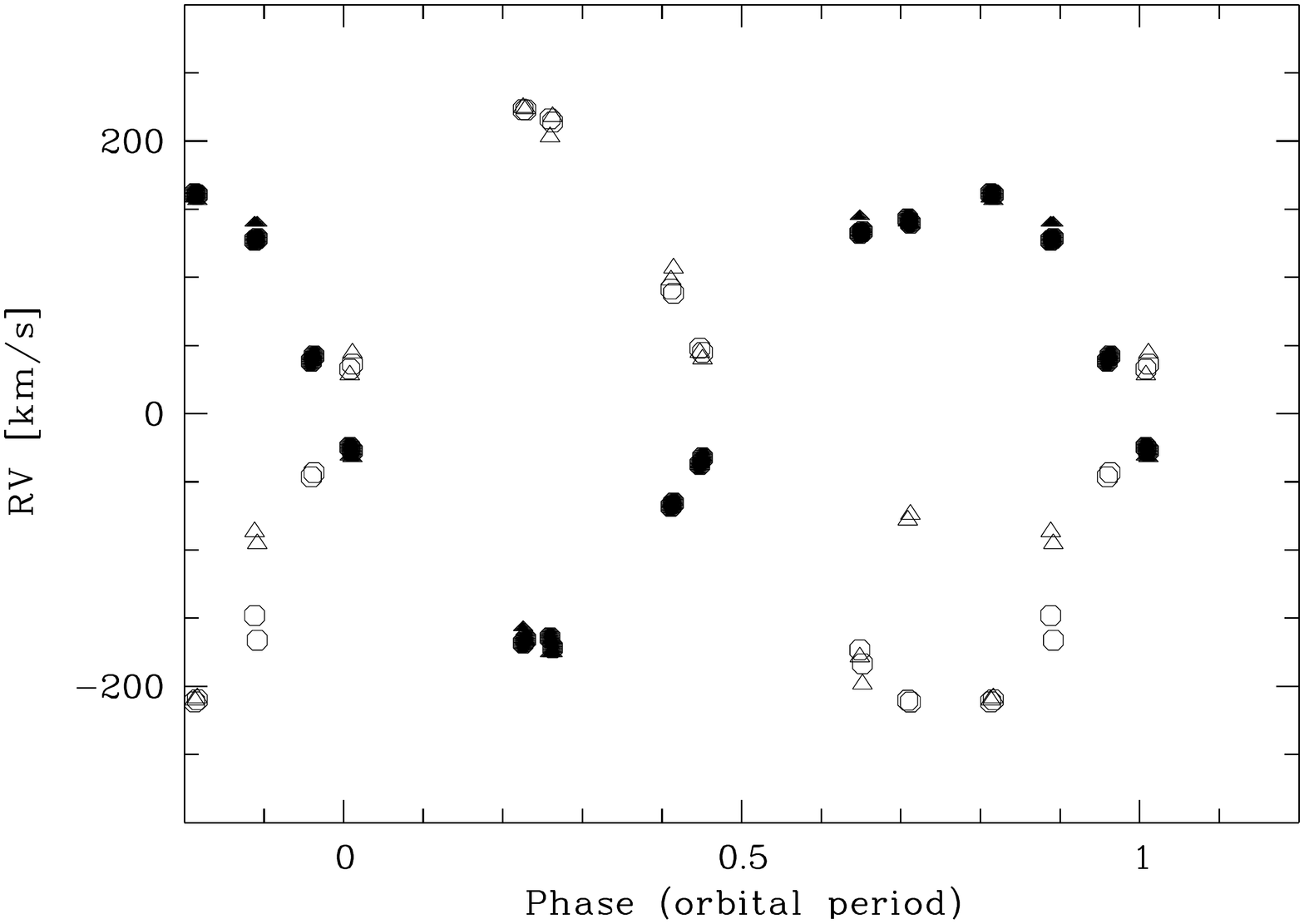}
\caption{H$\beta$ line ($\circ$) and Si{\sc iii}\,4553, 4568, 4574 lines
  ($\triangle$) radial velocities obtained after Fourier spectral decomposition
  (filled symbols star\,\#1, open symbols star\,\#2). 
 The two outliers in the secondary's Si{\sc iii} RVs occur when
  the spectra are of somewhat lower quality in that region. In other cases,
  when only one type of symbol is visible this is due to agreement resulting
  in the two symbols to be plotted above each other. The entire curve is
  shifted by the $\gamma$ velocity (227 $\pm$ 3 km\,s$^{-1}$) and centered around 0\,km\,s$^{-1}$.}
  \label{fig_korel}
\end{figure}


The corresponding velocities show that $P_{2}= 5.366$\,d is clearly a
binary orbital period and suggests a near circular orbit
(Fig.~\ref{fig_korel}).  The period $P_{\rm res}$ = 3.577\,d is actually not
present in the RVs. The fit for H$\beta$ RVs gives $K_{1} = 169 \pm 10$
km\,s$^{-1}$ and $K_{2} = 224 \pm 10$ km\,s$^{-1}$, i.e.\ a mass ratio $q$ of
about 0.75.

The analysis gives a small non-zero eccentricity of about $e = 0.08 $
with a longitude for the periastron of $\omega = 93$ degree.  Similar
figures were obtained by using the strongest Si{\sc iii} lines, providing
extra confidence on the above results.  { While this small eccentricity is
  compatible with $e = 0.00$ plus non-zero noise, we note the system is
  relatively young, so a not fully completed circularization is entirely
  possible.}  Residual radial velocities, corrected for orbital motion with
these parameters, are not sorted with $P_{1}$, i.e.\ that period does
not reflect in the orbital dynamics of the binary system. The entire set of
derived parameters is given in Table~\ref{tab_korelres}.

We calculated independent system parameters with {\rm PHOEBE} \citep{2011ascl.soft06002P} using 
the disentangled orbital light curve and assuming $T_{1}$ = $T_{2}$ = 27500 $K$, $P$ = 5.36654 d, 
$q$ = 0.75, 
$e$ = 0 and no third light.
Also we  assumed a detached configuration and synchronous rotation
of both components.
We find $a$  = 47 $R_{\odot}$, $i$ =  60.3 degree, 
stellar potentials 3.95 and  4.15 for the gainer and donor, respectively.
This solution implies $M_1$ = 27.7 $M_{\odot}$, $M_2$ = 20.8 $M_{\odot}$, $R_1$ = 15 $R_{\odot}$, $R_2$ = 11.8 $R_{\odot}$. These photometrically derived masses and orbital separation are consistent with the spectroscopic values  given in Table 4 for an inclination angle of $\approx$ 62 degree. 

\begin{figure*}
\includegraphics[viewport=55 41 572 775,width=0.33\textwidth]{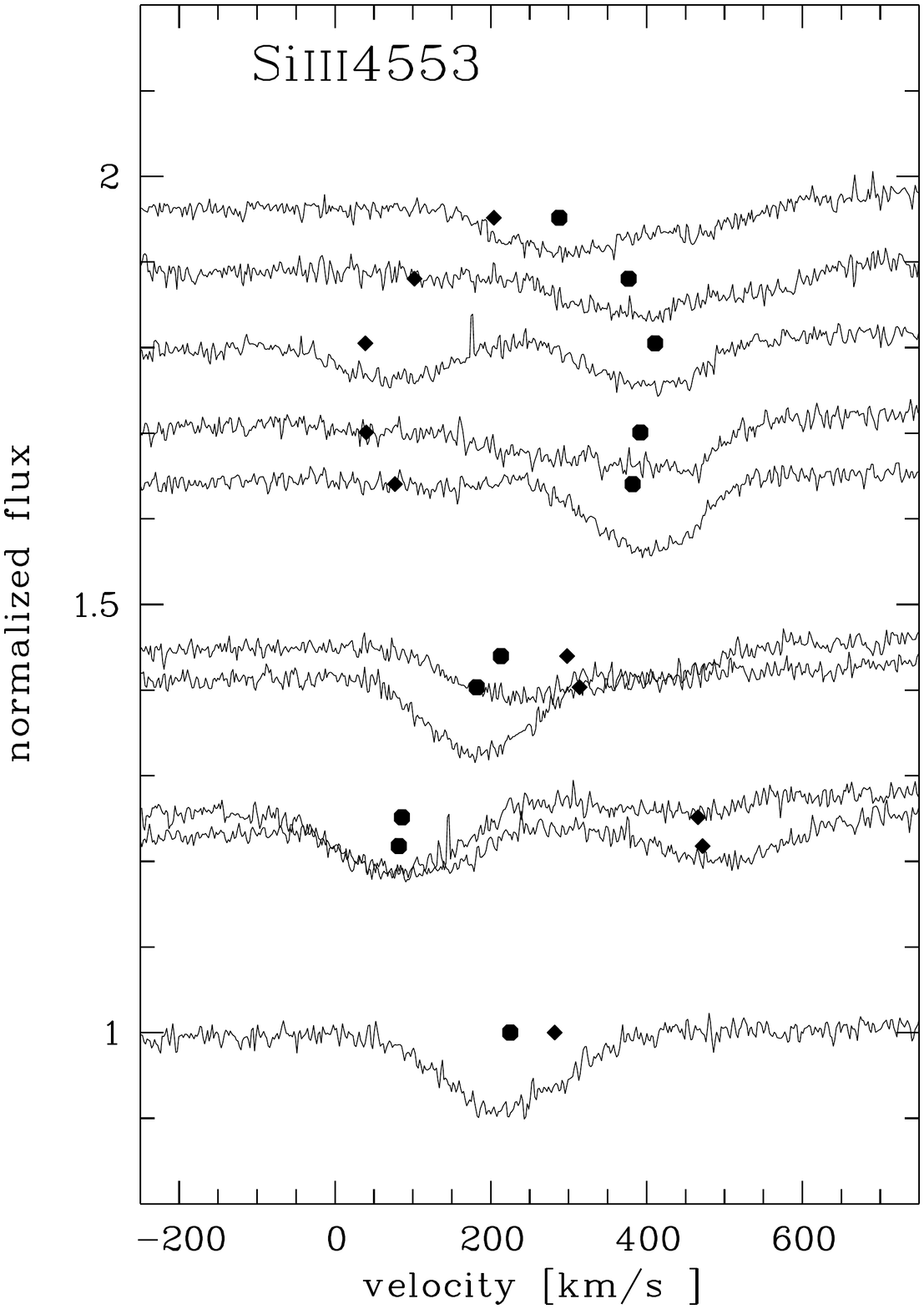}%
\includegraphics[viewport=55 41 572 775,width=0.33\textwidth]{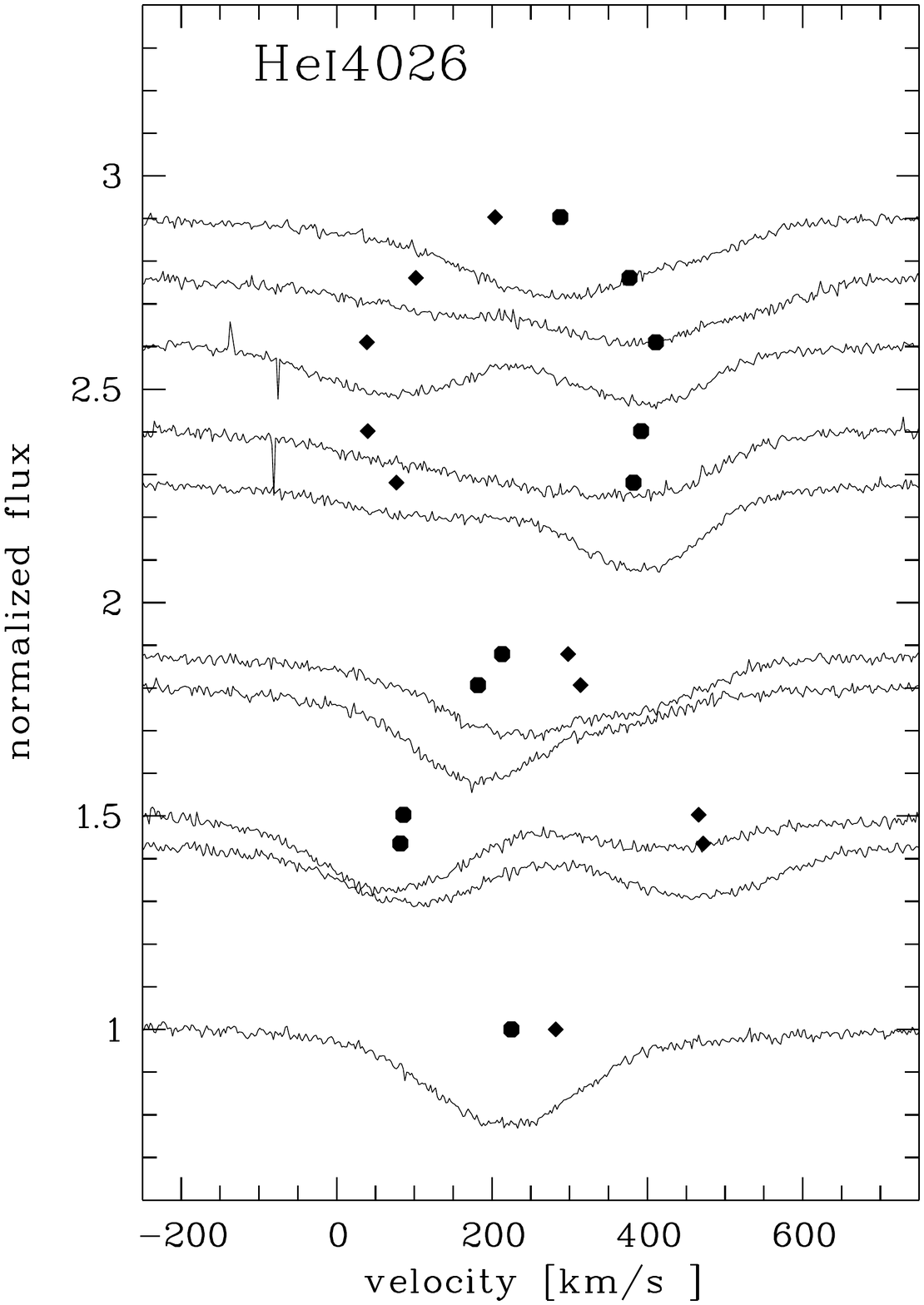}%
\includegraphics[viewport=55 41 572 775,width=0.33\textwidth]{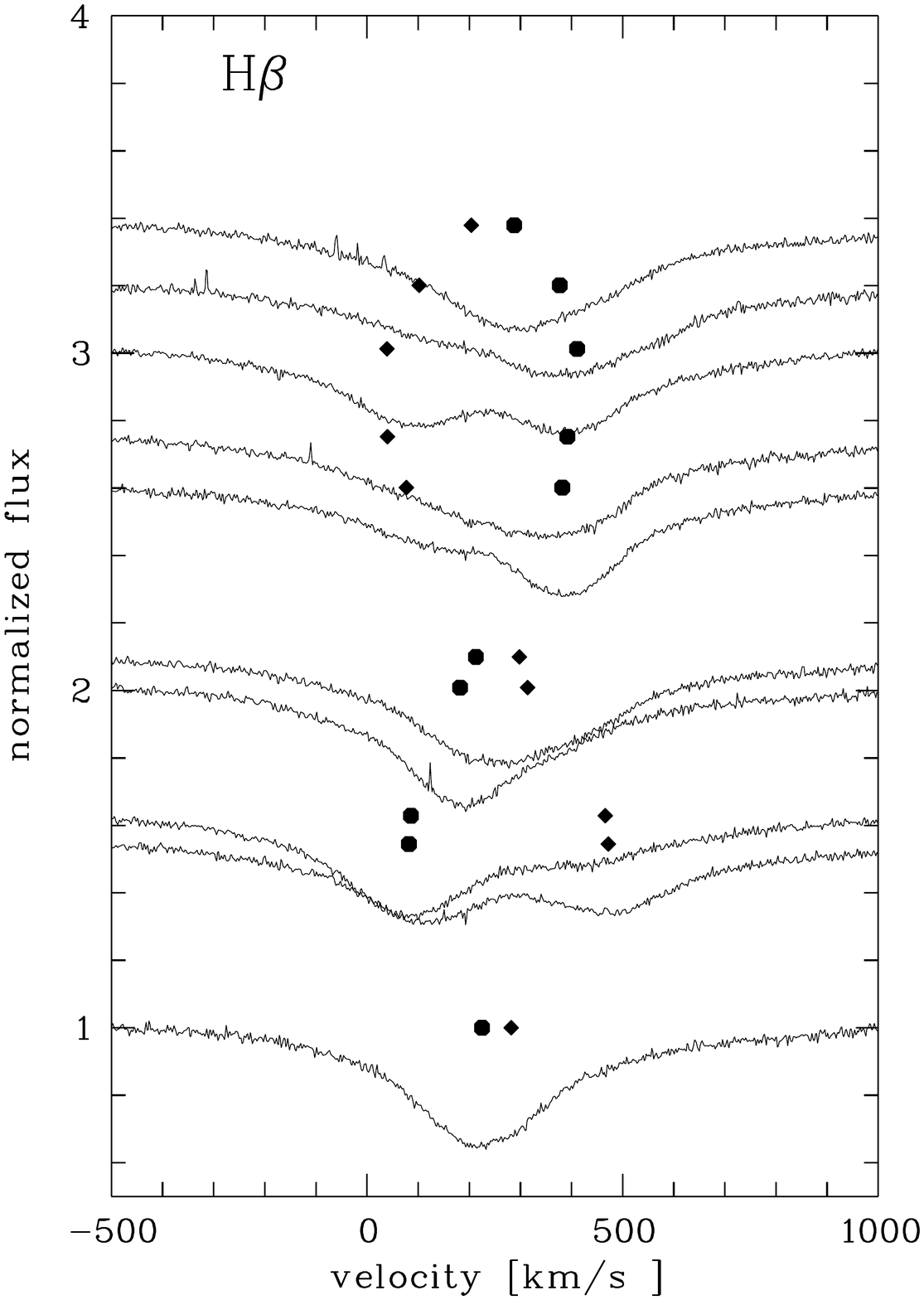}%
\caption{The spectral variability of Si{\sc iii}\,4553, He{\sc i}\,4026 and
  H$\beta$, sorted with the orbital period, from bottom to top. Offset is
  proportional to phase and according to values given in Table 1. The RVs as measured in H$\beta$ are shown as filled
  disks for star\,\#1 and diamonds for star\,\#2, respectively. 
}
  \label{fig_lpv}
\end{figure*}

The equivalent
widths of lines of both stellar components are highly variable and roughly anticorrelated.
The sum of both does vary as well, but with much lower
amplitude. Unfortunately, with the available data it cannot be decided whether
$P_{1}$ or $P_{2}$ does sort the data better.
Similarly, the line widths are variable (see Fig.~\ref{fig_lpv}),
with no clear preference for one of the two periods.

\section{Discussion}

\citet{2008IBVS.5868....1O} originally suggested a quadruple star in
(Aa+Ab)+(Ba+Bb) configuration to explain the light curve of BI\,108. 
Similarly, the light curve of CzeV343 was modeled as a double eclipsing binary with periods near a 3:2  ratio \citep{2012A&A...544L...3C}.
These authors speculate that this resonance might be the result of Kozai cycles and tidal friction
during the evolution of a quadruple system, but recognize the lack of theoretical work to confirm this conjecture and
that a simple coincidence cannot be discarded. In the case of  BI\,108,
the proximity to perfect resonance is 0.008\% (compared with 0.1\% for CzeV343) but the
circularity of the orbit is in principle inconsistent with Kozai cycles; they tend to increase the ellipticity of central binaries in multiple systems.  

For BI\,108
the absence of color variations would require the Ba+Bb system has equal components.
  To make Ba+Bb as dim as
  possible (in order to hide eventual spectral features), one has to demand
  that the eclipse in the light-curve is total, such that the entire Ba+Bb
  subsystem contributes about 20\% to the total flux as indicated by the
  photometric variability.  Under such assumptions the individual
  magnitudes would be 14.3 for Aa and Ab each, and 15.8 for Ba and Bb each,
  i.e. the subsystem A providing about 80\% of the total light in the $I$-band
  light curve. 
  Rapid
  rotation (beyond corotation of the order of 200\,km\,s$^{-1}$) for both
  components of Ba+Bb would still be required in order to hide spectral
  features from this pair from detection in the high $S/N$ data. Taking all
  the above together, such a quadruple  system cannot be ruled out and
  in spite of the lack of theoretical understanding, this will still turn out to be the hypothesis matching the requirements of Occam's razor.



Another possible multiple system would consist of three stars in
non-hierarchical orbits A+B+C, where $P_{1}$ is the result of
eclipses of the brighter component A by both B and C.  Indeed some
stable solutions exist for the unrestricted three body problem, for instance
those found by Lagrange where three bodies orbit at the vertices of a rotating
equilateral triangle.
However, this would not only require a parameter fine-tuning
beyond belief, but probably would still not be stable in a real stellar system
undergoing stellar evolution.  

 
If there would be only two stars, they would show variability  somehow resonantly synchronized with the
orbit. 
Whatever this mechanism is, to achieve it, the two components must
exert force on each other. This could be done through magnetic
fields  modifying the chemical element distribution on the stellar surfaces, hence producing line and continuum flux variability \citep{2007A&A...470.1089K}. 
Historically, it has been assumed that magnetic fields in OB stars are very rare if not absent.  However, according to \citet{2011IAUS..272..106P},
``... the increasing discoveries of fields in early B-type stars on the main sequence and pre-main sequence and in both young and evolved O-type stars show convincingly that  fossil fields can and do exist in stars with masses as large as 45 solar masses``.  However, 
in BI\,108 magnetic fields can be excluded with high probability,
 due to the missing signatures of chemical inhomogeneities in the spectral lines. 

It has been shown that excitation of
resonant non-radial $g$-mode pulsations should be a common phenomenon during
tidal evolution of eccentric double main sequence binaries and it is also an
effective mechanism for orbital decay
\citep{1999A&A...350..129W,2001A&A...366..840W}.  
Although this is a promising mechanism for
explaining the $P_{1}$ variability in principle, it is difficult to imagine how it could generate the
amplitude observed in the $P_{1}$ light curve.
This would require strongly perturbed stars, the visible stellar surface in the system varying by as much as 20\,\% between maximum and
minimum. 
Thus, other than for the double binary hypothesis, there is a theoretical framework for the variability, however the observed strength of the variation goes far beyond what has been explored in this framework. For this reason we  consider it a less likely option than the double binary hypothesis, but note that neither can it be  firmly ruled out.




\section{Conclusions}

We have presented the first spectroscopic study of the LMC system BI\,108
showing an unique 3:2 resonant photometric variability.  The immediate goal of
the study was to determine which of the two periods (if
not both) was due to an eclipsing binary. From the shape of the light-curves,
it was originally expected that $P_{1}$ would turn out to be the binary
period.  Surprisingly the radial velocities point to a SB2E system
orbiting with orbital period $P_{2}$ instead.

The mechanism responsible for $P_{1}$ remains unknown, but with the
present data we exclude a system with three components and magnetic interaction between two binary components. The 3:2 resonant tidal bulge 
would require amplitudes beyond what can realistically be expected from such a mechanism. 
 The remaining suggestion, a double binary system seen by chance (or eventually by still unknown dynamical reasons)  with a period ratio near 3:2, remains as the most plausible hypothesis.


\section*{Acknowledgments}
We acknowledge an anonymous referee for helping to improve a first version of this manuscript. ZK acknowledges support by Polish NCN grant 2011/03/B/ST9/02667. 
REM acknowledges support by Fondecyt grant 1110347 and  BASAL
PFB--06/2007.
We acknowledge Dr. Moffat for useful comments on a first version of this manuscript.
We also thank Igor Soszynski for reading the manuscript and Daniela Barr\'{\i}a for running {\rm PHOEBE}.


\bibliographystyle{mn2e} \bibliography{BI108}

\newcommand{\oneletter}[1]{#1}
\begin{thebibliography}{}

\bibitem[\protect\citeauthoryear{{Bayo}, {Rodrigo}, {Barrado Y Navascu{\'e}s},
  {Solano}, {Guti{\'e}rrez}, {Morales-Calder{\'o}n} \& {Allard}}{{Bayo}
  et~al.}{2008}]{2008A&A...492..277B}
{Bayo} A.,  {Rodrigo} C.,  {Barrado Y Navascu{\'e}s} D.,  {Solano} E.,
  {Guti{\'e}rrez} R.,  {Morales-Calder{\'o}n} M.,    {Allard} F.,  2008, \aap,
  492, 277

\bibitem[\protect\citeauthoryear{{Caga{\v s}} \& {Pejcha}}{{Caga{\v s}} \&
  {Pejcha}}{2012}]{2012A&A...544L...3C}
{Caga{\v s}} P.,  {Pejcha} O.,  2012, \aap, 544, L3

\bibitem[\protect\citeauthoryear{{Fitzgerald}}{{Fitzgerald}}{1970}]{1970A&A...%
..4..234F}
{Fitzgerald} M.~P.,  1970, \aap, 4, 234

\bibitem[\protect\citeauthoryear{{Hadrava}}{{Hadrava}}{1995}]{1995A&AS..114..3%
93H}
{Hadrava} P.,  1995, \aaps, 114, 393

\bibitem[\protect\citeauthoryear{{Ko{\l}aczkowski}, {Mennickent} \&
  {Rivinius}}{{Ko{\l}aczkowski} et~al.}{2010}]{2010ASPC..435..403K}
{Ko{\l}aczkowski} Z.,  {Mennickent} R.,    {Rivinius} T.,  2010, in {Pr{\v s}a}
  A.,  {Zejda} M.,  eds, Binaries - Key to Comprehension of the Universe
  Vol.~435 of Astronomical Society of the Pacific Conference Series, {The Study
  of Resonant Variability Observed in the Massive LMC System BI 108}.
p.~403

\bibitem[\protect\citeauthoryear{{Koornneef}}{{Koornneef}}{1983}]{1983A&A...12%
8...84K}
{Koornneef} J.,  1983, \aap, 128, 84

\bibitem[\protect\citeauthoryear{{Krti{\v c}ka}, {Mikul{\'a}{\v s}ek}, {Zverko}
  \& {{\v Z}i{\v z}{\'n}ovsk{\'y}}}{{Krti{\v c}ka}
  et~al.}{2007}]{2007A&A...470.1089K}
{Krti{\v c}ka} J.,  {Mikul{\'a}{\v s}ek} Z.,  {Zverko} J.,    {{\v Z}i{\v
  z}{\'n}ovsk{\'y}} J.,  2007, \aap, 470, 1089

\bibitem[\protect\citeauthoryear{{Lanz} \& {Hubeny}}{{Lanz} \&
  {Hubeny}}{2003}]{2003ApJS..146..417L}
{Lanz} T.,  {Hubeny} I.,  2003, \apjs, 146, 417

\bibitem[\protect\citeauthoryear{{Lanz} \& {Hubeny}}{{Lanz} \&
  {Hubeny}}{2007}]{2007ApJS..169...83L}
{Lanz} T.,  {Hubeny} I.,  2007, \apjs, 169, 83

\bibitem[\protect\citeauthoryear{{Maeder} \& {Meynet}}{{Maeder} \&
  {Meynet}}{2001}]{2001A&A...373..555M}
{Maeder} A.,  {Meynet} G.,  2001, \aap, 373, 555

\bibitem[\protect\citeauthoryear{{Ofir}}{{Ofir}}{2008}]{2008IBVS.5868....1O}
{Ofir} A.,  2008, Information Bulletin on Variable Stars, 5868, 1

\bibitem[\protect\citeauthoryear{{Petit}}{{Petit}}{2011}]{2011IAUS..272..106P}
{Petit} V.,  2011, in {Neiner} C.,  {Wade} G.,  {Meynet} G.,   {Peters} G.,
  eds, IAU Symposium Vol.~272 of IAU Symposium, {Observations of magnetic
  fields in hot stars}.
pp 106--117

\bibitem[\protect\citeauthoryear{{Pietrzy{\'n}ski}, {Thompson}, {Graczyk},
  {Gieren}, {Udalski}, {Szewczyk}, {Minniti}, {Ko{\l}aczkowski}, {Bresolin} \&
  {Kudritzki}}{{Pietrzy{\'n}ski} et~al.}{2009}]{2009ApJ...697..862P}
{Pietrzy{\'n}ski} G.,  {Thompson} I.~B.,  {Graczyk} D.,  {Gieren} W.,
  {Udalski} A.,  {Szewczyk} O.,  {Minniti} D.,  {Ko{\l}aczkowski} Z.,
  {Bresolin} F.,    {Kudritzki} R.,  2009, \apj, 697, 862

\bibitem[\protect\citeauthoryear{{Pietrzynski} \& {Udalski}}{{Pietrzynski} \&
  {Udalski}}{2000}]{2000AcA....50..337P}
{Pietrzynski} G.,  {Udalski} A.,  2000, \actaa, 50, 337

\bibitem[\protect\citeauthoryear{{Prsa}, {Matijevic}, {Latkovic}, {Vilardell}
  \& {Wils}}{{Prsa} et~al.}{2011}]{2011ascl.soft06002P}
{Prsa} A.,  {Matijevic} G.,  {Latkovic} O.,  {Vilardell} F.,    {Wils} P.,
  2011, Astrophysics Source Code Library, p.~6002

\bibitem[\protect\citeauthoryear{{Rivinius}, {Mennickent} \&
  {Ko{\l}aczkowski}}{{Rivinius} et~al.}{2011}]{2011IAUS..272..541R}
{Rivinius} T.,  {Mennickent} R.~E.,    {Ko{\l}aczkowski} Z.,  2011, in {Neiner}
  C.,  {Wade} G.,  {Meynet} G.,   {Peters} G.,  eds, IAU Symposium Vol.~272 of
  IAU Symposium, {The resonant B1II + B1II binary BI 108}.
pp 541--542

\bibitem[\protect\citeauthoryear{{Schlegel}, {Finkbeiner} \&
  {Davis}}{{Schlegel} et~al.}{1998}]{1998ApJ...500..525S}
{Schlegel} D.~J.,  {Finkbeiner} D.~P.,    {Davis} M.,  1998, \apj, 500, 525

\bibitem[\protect\citeauthoryear{{Schmidt-Kaler}, {Gochermann}, {Oestreicher},
  {Grothues}, {Tappert}, {Zaum}, {Bergh{\"o}fer} \& {Brugger}}{{Schmidt-Kaler}
  et~al.}{1999}]{1999MNRAS.306..279S}
{Schmidt-Kaler} T.,  {Gochermann} J.,  {Oestreicher} M.~O.,  {Grothues} H.,
  {Tappert} C.,  {Zaum} A.,  {Bergh{\"o}fer} T.,    {Brugger} H.~R.,  1999,
  \mnras, 306, 279

\bibitem[\protect\citeauthoryear{{Szymanski}}{{Szymanski}}{2005}]{2005AcA....5%
5...43S}
{Szymanski} M.~K.,  2005, \actaa, 55, 43

\bibitem[\protect\citeauthoryear{{Udalski}, {Szymanski}, {Soszynski} \&
  {Poleski}}{{Udalski} et~al.}{2008}]{2008AcA....58...69U}
{Udalski} A.,  {Szymanski} M.~K.,  {Soszynski} I.,    {Poleski} R.,  2008,
  \actaa, 58, 69

\bibitem[\protect\citeauthoryear{{{\v S}koda} \& {Hadrava}}{{{\v S}koda} \&
  {Hadrava}}{2010}]{2010ASPC..435...71S}
{{\v S}koda} P.,  {Hadrava} P.,  2010, in {Pr{\v s}a} A.,  {Zejda} M.,  eds,
  Binaries - Key to Comprehension of the Universe Vol.~435 of Astronomical
  Society of the Pacific Conference Series, {Fourier Disentangling Using the
  Technology of Virtual Observatory}.
p.~71

\bibitem[\protect\citeauthoryear{{Witte} \& {Savonije}}{{Witte} \&
  {Savonije}}{1999}]{1999A&A...350..129W}
{Witte} M.~G.,  {Savonije} G.~J.,  1999, \aap, 350, 129

\bibitem[\protect\citeauthoryear{{Witte} \& {Savonije}}{{Witte} \&
  {Savonije}}{2001}]{2001A&A...366..840W}
{Witte} M.~G.,  {Savonije} G.~J.,  2001, \aap, 366, 840

\end{thebibliography}

\bsp 
\label{lastpage}
\end{document}